# Using Socrative and Smartphones for the support of collaborative learning

Mohammad Awedh, Ahmed Mueen,Bassam Zafar, Umar Manzoor
King Abdulaziz University, Saudi Arabia, Jeddah

## Abstract

*The integration of new technologies in the classrooms opens new possibilities for the teaching and learning process. Technologies such as student response system (e.g. Clicker) are getting popularity among teachers due to its effects on student learning performance. In this study, our primary objective is to investigate the effect of Socrative with combination of smartphones on student learning performance. We also observed the benefits of interactivity between the teacher and the students and among classmates, which positively influences collaborative learning and engagement of students in the class. We test these relationships experimentally in a community college class environment using data from a survey answered by students in information technology associate degree. The results of our study reveal that collaborative learning and engagement of student in the class improves student learning performance. We highly recommend these tools in educational settings to support the learning process.*

## 1. Introduction

Computers and related technology become essential part of a classroom to enhance 21[st]century teaching and learning experiences(Luu & Freeman, 2011; Windschitl, 2009).Information technology provides educational institutions an exceptional opportunity to increase student enthusiasm and enhance learning outcomes (Roblyer & Wiencke, 2003). The technology tools that have been used by educational institutions in current years include popular course management systems such as Web CT and Blackboard. They normally provide tools for delivery of content, quizzing, and file sharing(Boettcher, 2003; Oliver, 2001).Beyond course management tools, web annotation software permits individuals to evaluate and integrate their ideas with present online content such as the emergence of Web with blog tools for students to communicate their ideas and receive feedback in the form of comments(Richardson, 2006). A variety of web tools are also available to help students to use problem-based approach to learning from online resource.  The Intel (2006)Showing Evidence Tool provides a scaffold to support students as they create a claim and then support or refute it with suitable evidence. The latest among them are tools like Weebly, Edmodo, Class Dojo, etc. There are some hardware related tools such as clickers. It is interactive remote response devices that transmit and record student responses to questions providing immediate feedback about the learning process (Homme et al, 2004).In this study, we focus on Socrative,which is an online student response system that allows teachers to effortlessly generate quizzes and other educational exercises for their students and monitor their students' response and progress in real time.

Previous studies show interest in the role of clickers (Blasco et al., 2012) where teacher generates a question and shows it on the projector, while students use the clickers to choose one of the answers. It requires to purchase devices to record student responses. Whereas, Socrative only needs usual resources today like Internet and smart phone(Matthew & Anne, 2012). Furthermore, Socrative allow teachers to design different activities and control the flow of quizzes. The students' responses reports can be view online as a Google spreadsheet or as an excel file. Mostly, researcher analyzes clickers as individual student tool other than collaborative tool (Fies and Marshall, 2006). In this study, we investigated Socrative as collaborative learning tool. In addition, prior research has been carried out with teachers prospective(Méndez and Slisko, 2013). We have examined Socrative with students who possibly have larger conceptual gain. Addressing these issues, our primary objective is to investigate the impact of Socrative on student learning performance. We propose that interaction between the teacher and the students and among students using Socrative affects student collaborative learning and enhance student learning experiences.

## 2. Collaborative learning

In recent years special attention has been devoted to the tools that facilitate collaborative learning in educational institutions (Fischer et al., 2007; Hernández-Leo et al., 2006).Collaborative learning is a learning that contains sharing knowledge and experiences, in which students teach and learn from each other and develop interdependence(Panitz, 1996).Through the process of collaboration in a collaborative learning, students are able to efficiently obtain huge amount of information, which is useful to student in generating new ideas for effective learning (Lipponen, 2002). It gives student ability to think critically(Angeli et al., 2003) and encourages students to contribute in giving the answer and expressing their opinion (Lantz,2010). Consequently, students become active learner in their learning process and collaborate in the construction of their own knowledge. The collaborative learning method allows students to have deeper understanding of the subject matter and helps student to link new information with previous knowledge (Kennedy & Cuts, 2005).Collaborative learning is an essential part of active learning. Active learning is defined as conscious effort by a teacher to excite his student to participate explicitly in a classroom. It is an exercise including techniques that involve students in the learning process where students do more than inactively listen to a lecture. Studies showed that students learn better when they participate in active learning process (Mayer & Wittrock, 2006), their academic performance increase (Yoder & Hochevar, 2005), and they do well in exams (Knight &Wood, 2005) over traditional learning process. Furthermore, combining active collaborative learning with technology enhanced student academic performance. The researchers observed that students who are skilled technology users are more active learner as compare to other students. Therefore, many researchers have been attracted to use technology with collaborative learning (Kreijns et al.,2003).

In this study, we used Socrative tool to improve the efficiency of the active collaborative learning. Socrative allow students to cognitively process questions asked by the teacher and to increase participation. Teachers using Socrative need to bring important changes in their class format. They have to encourage students to discuss ideas, give opinion, debate point of view critically. Socrative facilitate students to be the part of knowledge creation, so that students sense that they are participating in their own learning. We consider that Socrative increases the degree of collaboration learning gained by the students during the learning process, which enhances student overall performance.

## 3. Smartphones

With the rapid growth of information and communication technologies a special importance is given to mobile learning as a new trend of development based on interconnection of devices. The Mobile learning is a type of e-learning, a technique for distance education using computer and Internet technology, which offers learning through wireless handheld devices like smart phones and tablets. (Georgievand Smrikarov, 2004). Integration of smartphones in the class helps to enhance individual and group learning outcomes along with enabling more interactive discussions among group members (Duncan et al., 2012). Smartphones are very flexible;students can use them anytime, are always with student, and are always on (Kolb, 2011)**.**The use of smart phones tools can be useful for the teachers because they can control the students' learning in real time (Manuguerra&Petocz, 2011).The aim of our study is to investigate the involvement of higher education students in technology and the effect of collaboration learning in their academic performance.  Several educational institutions are using clicker an electronic device to get the feedback or answers of students in real time. Student use these devices to answer any question projected on the projector by their teacher(Caldwell, 2007).Teacher question must be of multiple-choice type and he has to purchase clicker devices to distribute among the students. In contrast, today there are some websites available which provide posting questions and receiving answers services to the teachers.   Some of them are commercial websites like Poll Everywhere(www.polleverywhere.com) and Go Soapbox (www.gosoapbox.com). Whereas, Socrative is free (www.socrative.com) (Matthew, 2012), which only requires internet anda smartphone.Therefore, we decided to integrate Smartphone and Socrativein the class for the following benefits:

- Learning using Socrative encourages students in both independent and collaborative learning experiences.
- Collaborating with classmates as a result of using Socrative increases students' engagement.
- Interactivity with the teacher as a result of using Socrative increases students' collaborative learning.

## 4. Method

### 4.1 Participants

The research was conducted at community college in Jeddah, Saudi Arabia. The samples were taken from 38 students; they were attending computer architecture course.They were in their third semester. They attended classes 2 days a week for 4 hours. In total, two classes participated.Their ages ranged from 18 to 24. There is no female student in the college. Therefore, all the samples were taken from male students.

## 4.2 Procedure

The teachers were to standardized their course material (e.g. lectures and PowerPoint slides) to ensure that two classes covered the same material in a similar way. They were requested to ynchronize their method of delivering lectures and teaching techniques in the classrooms(Yourstone et al., 2008). Inone semester, each of the two classes was given 6 multiple-choice quizzes, with 10 questions per quiz, student use Socrative to answer all questions. The marks for each quiz were 5% from their total course marks. The arrangement of all sessions was always the same. In class, the teacher explains the topic for approximately twenty minutes. Then he poses 5 questions related to the topic on Socrative. Students were supposed to answer them individually within ten minutes. The teacher collected all the answers of the students and pinpoint questions in which there is a major difference of opinions. Later, teacher foam groups of three or four students who answered different. The students discuss their answers for twenty minutes in the groups. Teacher then asked students to do the quiz again individually using Socrative. This time student answered according to the discussion they made in the groups. Fig. 1 shows the experimental design of this study. Afterwards, teacher asked another round of 5 questions with similar procedure.

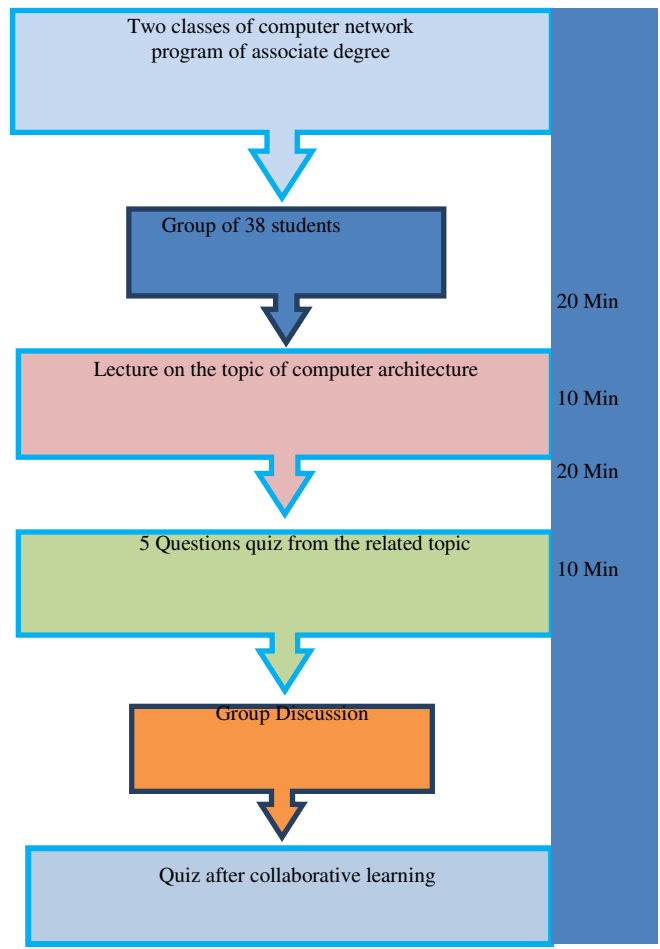

Figure 1: Experimental design of the study

**4.3 Data collection and measures**

After completion of the experiment with the Socrative in class, the students were asked to give their judgment about it. A questionnaire consisting of 20 items was designed. The survey was conducted at the end of the semester. Each participant was provided with a questionnaire and a brief background to the study. The survey contained questions about the student's impressions of Socrative and the advantages of using this technology. There were five-point Likertscale items were used that ranged from 1 (strongly disagree) to 5 (strongly agree)(Blasco et al., 2012). The survey also included questions about biographical information.

# 5. Results

## 5.1 Experience of students with Socrative

The survey result (Table 1) showed that student feels that collaborative learning significantly affect student learning performance. Collaborative learning allows students to exchange information with classmates, and make students more excited. The students stated that these experiences have assisted them to be more active in the classes, help them to understand concepts, facilitate to work in groups and understand their level of knowledge.

## 5.2. Observation of the students' interaction with the applications

According to the informal data gathered from the researcher's observations, students seemed enthusiastic with the use of Socrative in class. Their level of excitement did not reduce for the entire experiment time frame. They showed high level of engagement during group discussions. They feel very interesting to do quiz using mobile. When asking informally with few students about their experience using Socrative in class.

Table 1: Answersof the students to the questions

| Questions | strongly agree | Agree | neutral | disagree | strongly disagree |
|---|---|---|---|---|---|
| Gives me the opportunity to discuss with classmates | 32% | 63% | 5% | 0% | 0% |
| Allows the exchange of information with classmates | 35% | 59% | 6% | 0% | 0% |
| Gives me the opportunity to discuss with the teacher | 21% | 68% | 9% | 2% | 0% |
| Allows the exchange of information with the teacher | 18% | 72% | 8% | 2% | 0% |
| I felt that I actively collaborated in my learning experience | 11% | 73% | 12% | 4% | 0% |
| I felt that I had freedom to participate in my own learning experience | 10% | 75% | 12% | 3% | 0% |
| In this method, my classmates and faculty interactions made me feel valuable. | 3% | 73% | 18% | 5% | 1% |
| This method has favored my personal relationships with my classmates and teacher. | 30% | 55% | 10% | 5% | 0% |
| This can improved my comprehension of the concepts studied in class | 25% | 66% | 8% | 1% | 0% |
| This method can lead to a better learning experience | 27% | 67% | 5% | 1% | 0% |

They said it helped to understand concepts; it facilitated the interaction with the classmates and teacher; and it helped them to be motivated. It seems that the structure of the Socrative website was easy to understand and the quiz was easy to answer. Furthermore, students were seems to be comfortable with number of questions given and time assigned to answer the questions or discuss the questions with classmates.

## 6. Discussion and conclusion

Student Response Systems (SRS) transform any classroom into an active learning environment. As more and more educational institutions integrate student response system into their classes to enhance the learning process, it becomes essential to have a thorough understanding of these systems and to know what types of SRS are available. In spite the popularity of clickers there are many new applications arising in internet. Socrative is one of them and is a very useful tool because it helps teachers to monitor learning of all the students in real time. In addition, teachers are not required to invest money to buy the devices such as clickers.In this study, our primary objective was to identify that Socrative and smart phones are suitable tools that canfacilitate active learning in classroom. This result also suggests that students perceive that Socrative supports the learning and increases the student motivation. In addition, it helped them to be aware of their level of knowledge and facilitates the understanding of the concepts and significantly increases their learning process. These tools also increase student level of communication with their classmates and teachers and support collaborative information exchange among them. It develops communication skills and a collaborative spirit among students and this process helps them improve their learning performance. Furthermore, students feel that their answers and opinions are given value by the teacher and their classmates. Simultaneously, it easy for teachers to check how many students understands the concept.A limitation of this study is that we collected sample of those students who have used Socrative whereas data from control group of non-user were missing. Therefore, further research would be to test two different student groups: Socrative users and non-users. We strongly recommend the use of Socrative in the class as a tool to enhance the learning experience. We conclude that Socrative improve students level of interactivity, which helps student to be active in class and have collaborative learning, which also increases student engagement in the learning process.